\documentclass[9pt,conference]{IEEEtran}

\usepackage{bm} 

\usepackage{dcase2025,amsmath,graphicx,url,times,booktabs, tabularx}

\title{Acoustic Imaging for UAV Detection: Dense Beamformed Energy Maps and U-Net SELD}

\name{
  Belman Jahir Rodríguez$^{1}$\quad
  Sergio F.~Chevtchenko$^{1}$\quad
  Marcelo Herrera Martínez$^{2}$\quad
  Yeshwanth Bethi$^{1}$\quad
  Saeed Afshar$^{1}$
}

\address{
  $^{1}$\,International Centre for Neuromorphic Systems (ICNS), Western Sydney University, Australia\\
  $^{2}$\,Universidad de San Buenaventura, Colombia
}

\begin{document}

\maketitle

\begin{abstract}
We introduce a U-net model for 360° acoustic source localization formulated as a spherical semantic segmentation task. Rather than regressing discrete direction-of-arrival (DoA) angles, our model segments beamformed audio maps (azimuth × elevation) into regions of active sound presence. Using delay-and-sum (DAS) beamforming on a custom 24-microphone array, we generate signals aligned with drone GPS telemetry to create binary supervision masks. A modified U-Net, trained on frequency-domain representations of these maps, learns to identify spatially distributed source regions while addressing class imbalance via the Tversky loss. Because the network operates on beamformed energy maps, the approach is inherently array-independent and can adapt to different microphone configurations and can be transferred to different microphone configurations with minimal adaptation. The segmentation outputs are post-processed by computing centroids over activated regions, enabling robust DoA estimates. Our dataset includes real-world open-field recordings of a DJI Air 3 drone, synchronized with 360° video and flight logs across multiple dates and locations. Experimental results show that U-net generalizes across environments, providing improved angular precision, offering a new paradigm for dense spatial audio understanding beyond traditional Sound Source Localization (SSL). We additionally validate the same beamforming-plus-segmentation formulation on the DCASE 2019 TAU Spatial Sound Events benchmark, showing that the approach generalizes beyond drone acoustics to multiclass Sound Event Localization and Detection (SELD) scenarios.

\end{abstract}


\section{Introduction}
\label{sec:intro}

SSL is a fundamental task in spatial audio analysis, with applications ranging from surveillance, security, search and rescue, environmental monitoring, and wildlife tracking \cite{martinez-carranza_review_2020}. SSL is often reduced to estimating the DoA of sound sources. The DoA is usually defined as the azimuth and elevation angle of the direction of the audio source while ignoring the distance to it. Microphone arrays can be steered to act as spatial filters, which enables manipulation of the array's directivity (also referred to as beamforming).\cite{bai_acoustic_2013}. Traditional SSL techniques rely on signal processing algorithms such as time-difference of arrival (TDOA), generalized cross-correlation with phase transform (GCC-PHAT)\cite{chaudhary_sound_2014}, Steered-response power-phase transform (SRP-PHAT)\cite{grinstein_steered_2024} and multiple signal classification (MUSIC)\cite{schmidt_multiple_1986}, or beamforming methods like (DAS)\cite{perrot_so_2021}. While effective in controlled environments, these approaches degrade in performance under noise, reverberation, or when dealing with moving sources and complex acoustic scenes. Recent deep learning methods have achieved notable improvements in accuracy and robustness, even in challenging scenarios with noise, reverberation, and multiple simultaneous sources \cite{A_survey_Grumiaux}. In outdoor long-range scenarios, localization performance is frequently SNR-limited by propagation and non-stationary ambient noise. As distance increases, high-frequency UAV components attenuate rapidly and background noise becomes partially spatially coherent (e.g., wind, vegetation, traffic), reducing effective array gain. This motivates hybrid designs where a computationally simple beamformer produces an interpretable spatial map, and learning is used to denoise and regularize the map structure rather than operating directly on raw multichannel waveforms.
In recent years, the use of multichannel audio processing and visual perception has gained traction \cite{jekaterynczuk_survey_2023}, particularly with the use of microphone arrays and beamforming to create spatial energy maps \cite{wang_progress_2025} essentially turning sound into images, some of those systems are known as acoustic cameras.\cite{simeoni_deepwave_2019}.

The Acoustic "imaging" enables the use of powerful computer vision architectures, especially convolutional neural networks (CNNs), to perform spatial reasoning on sound scenes. However, most deep learning-based SSL models outputs direction-of-arrival (DOA) angles or coordinates, typically via classification or regression. Few have explored frame-based spatial segmentation of the full acoustic field analogous to semantic segmentation in images.

In this work, we propose a novel spherical segmentation framework for localizing sound sources using microphone arrays. Inspired by the image recognition paradigm in computer vision  like YOLO \cite{redmon_you_2016} or Deeplapv3\cite{chen_rethinking_2017},  we develop a U-Net-based architecture\cite{ronneberger_u-net_2015} that performs binary segmentation over a 2D spherical acoustic image (azimuth × elevation) derived from DAS beamforming. Instead of regressing point estimates of direction, our model learns to segment the acoustic field, highlighting regions in space where sound sources are present. This formulation enables spatial mapping of targets such as drones.
To enable supervised learning, we construct a labeled dataset based on
real-world recordings of a DJI Air 3 drone in open-field conditions on
different days and locations. The dataset is publicly
available. \cite{rodriguez_uav_2026} \footnote{\url{https://doi.org/10.26183/07nm-4p35}} It includes 24-channel audio, GPS-aligned binary masks and 360\textdegree{} video.

Our approach provides a scalable way to learn acoustic semantic segmentation, supporting generalization to other sound sources beyond drones. Potential applications include drone detection and tracking, acoustic camera visualization, multi-source scene understanding, and real-time sound field monitoring.

The main contributions are:
\begin{itemize}
    \item \textbf{Segmentation formulation for SSL/SELD:} we re-parameterize localization as dense spatial mask prediction over azimuth--elevation maps, enabling joint detection and localization via a single output representation.
    \item \textbf{Beamforming-to-image pipeline for 360$^\circ$ maps:} we build frequency-binned beamformed map tensors and introduce a polar reprojection for hemisphere coverage to better match convolutional spherical distortions.
    \item \textbf{Real open-field UAV dataset with GPS supervision:} we provide a field dataset with synchronized multichannel audio and drone telemetry for learning spatial segmentation under realistic outdoor recordings.
    \item \textbf{Cross-domain validation on TAU Spatial:} we re-cast DCASE 2019 Task~3 into multiclass spatial segmentation and demonstrate that the same formulation generalizes beyond drones.
\end{itemize}

\section{Related Work}
\label{sec:related}

Acoustic perception tasks
explicitly distinguish three sequential objectives: detection, classification, and localization. Many classical SSL systems tackle them in isolation rather than in an integrated pipeline \cite{martinez-carranza_review_2020}. Recent Sound Event Detection, Localisation and Classification(SELD/SELC) systems, also referred to as SSL, encompass two tasks: sound event detection (SED) and DoA, which are separate outputs of the neural network. The SED branch performs a multi-label classification task, and the DoA branch performs a multi-output regression task, as cited in \cite{santos_w2v-seld_2023} and \cite{adavanne_direction_2018}.

Our work resolves
SELD/SELC as a semantic-segmentation
problem; the resulting mask simultaneously answers: (i) is
there a source? (detection), (ii) where is it? (localisation) and through
class specific training (iii) what type is it? (classification). 

Conceptually, this paper keeps the SELD/SE objective but changes the output parameterization: instead of predicting class-wise event activity and DoA regression, we predict class-wise spatial occupancy masks over an azimuth--elevation grid. DoA estimates are then obtained by simple geometric post-processing (centroiding).

\subsection{Classical Sound‑Source Localisation (SSL)}
\label{sssec:Classic SSL}

In early SSL systems, DAS beamformer was the central processing block because it is algorithmically simple, computationally light, and easy to implement in hardware \cite{van_veen_beamforming_1988}. Later refinements improved robustness under real-world conditions: GCC-PHAT and SRP-PHAT introduce phase-based spectral weighting to combat reverberation \cite{knapp_generalized_1976,dibiase_high-accuracy_2000}, while the Minimum Variance Distortionless Response (MVDR) beamformer assigns microphone-specific weights that minimise noise without distorting the desired direction \cite{capon_high-resolution_1969}.

High‑resolution sub‑space methods go a step further.  
MUSIC \cite{schmidt_multiple_1986} and ESPRIT \cite{roy_esprit-estimation_1989} give sharper peaks, but assume narrow‑band, non‑coherent sources and perfect array calibration.   

In practice, classical methods do a good job at \textbf{localising} a single source in low‑noise scenes, and with thresholding they can give a basic \textbf{detection} flag.  
They do not, however, offer \textbf{classification}.  
Performance also drops when the number of active sources grows or when the room is highly reverberant \cite{localization_review_rascon}.  

\subsection{Deep Learning (DL) for SSL}

Large audio datasets and cheaper GPUs have made deep learning attractive for SSL \cite{martinez-carranza_review_2020}.  
Different deep learning architectures have been explored:

\begin{itemize}
  \item \textbf{CNNs} learn spatial–spectral patterns directly from spectrograms or MFCCs \cite{tan_sound_2021}.  
  \item \textbf{RNNs} and gated variants track moving sources by modelling temporal context \cite{adavanne_localization_2019}.  
  \item \textbf{Graph Neural Networks} capture the geometry of distributed microphone arrays \cite{bertrand_applications_2011}.  
  \item \textbf{Hybrid models} mix CNN encoders with RNN or Transformer decoders for stronger temporal cues \cite{wu_deep_2019, ji_transmusic_2024}.  
\item \textbf{U‑Net family}: Encoder–decoder U‑Nets have become popular because skip connections recover lost details in downsampling and give rich per-pixel output.  
Lee \textit{et al.} reach sub‑degree accuracy for overlapping sources by correcting the problems associated with DAS beam at low frequency and suppressing side‑lobes at high frequency \cite{lee_deep_2021}.  
Building on this idea, Zhou \textit{et al.} introduced audio‑visual segmentation (AVS) in which a TPAVI‑conditioned U‑Net injects audio cues at every scale to produce pixel‑wise masks of the \emph{visible} sounding objects \cite{zhou_audio-visual_2023}.  
Other works add a second head so that the same network performs sound‑event detection and localisation (SELD) simultaneously, while Dense‑U‑Net further extends the concept to dynamic, high‑noise audio‑visual scenes \cite{datta_dense-u-net_2024}.  
Unlike AVS, our approach removes the dependency on vision, thereby enabling efficacy even when the source lies outside the camera's field‑of‑view or under poor visibility.

\end{itemize}

Deep networks often produce false positives when trained exclusively on segments that contain sources. \cite{yalta_sound_2017} show that incorporating \emph{silence} frames into the training set improves robustness to background noise and prevents ghost detections. 
Inspired by this finding, we augment our drone dataset with “no‑drone’’ recordings, enabling the Tversky‑loss to direct the U‑Net to learn a calibrated decision boundary between the presence and absence of sound source of interest. 

\subsection{Gap Analysis and Motivation}

\textbf{Speed vs.\ Robustness.}  
Classical DAS‑based methods offer low computational latency but lose resolution and struggle with heavy noise or many sources.  
Fully trained DL models can be robust to noise and can detect, localize, and classify, but they usually require large training sets.

\textbf{Hybrid path.}  
By keeping a DAS front‑end and adding a light U‑Net back‑end we can:

- Keep end‑to‑end with low latency for real time applications.

- Learn to sharpen beams and suppress artifacts.

- Output detection, localization, and class labels in one shot — matching the three functional blocks proposed by \cite{martinez-carranza_review_2020}.

- A multiclass validation on DCASE 2019 TAU Spatial Sound Events, showing that the segmentation formulation remains competitive in DoA accuracy beyond drone data.

This mixed strategy directly addresses the open issues listed in recent surveys \cite{A_survey_Grumiaux, jekaterynczuk_survey_2023} and forms the basis of the method introduced in the present study.

\section{METHODOLOGY}
\label{sec:methodology}

\begin{figure*}[ht]
    \centering
    \includegraphics[width=\textwidth]{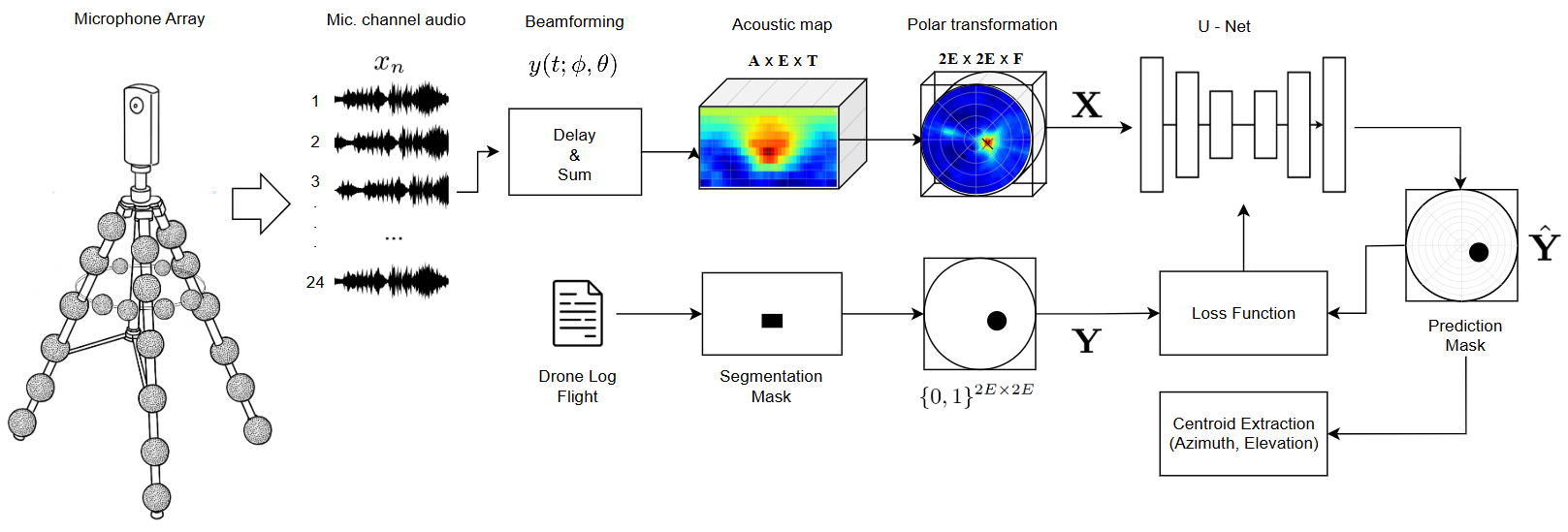} 
    \caption{The overall End-to-end pipeline: multichannel acquisition $\rightarrow$ DAS beamforming over azimuth/elevation $\rightarrow$ spectral binning $\rightarrow$ polar reprojection $\rightarrow$ U-Net segmentation $\rightarrow$ centroid-based DoA estimation.}
    \label{fig:Pipeline}
\end{figure*}


\subsection{System Overview}

\Cref{fig:Pipeline} shows an overview of the proposed system pipeline. It consists of a custom microphone array for capturing multichannel audio, DAS beamforming to generate spatial energy maps, dataset construction with GPS-based labeling, a U-Net segmentation model, and a centroid-based post-processing step for estimating the direction of arrival (DOA).

\subsection{Microphone Array Design and Recording Setup}
\begin{figure}[ht]
\centerline{\includegraphics[width=\columnwidth]{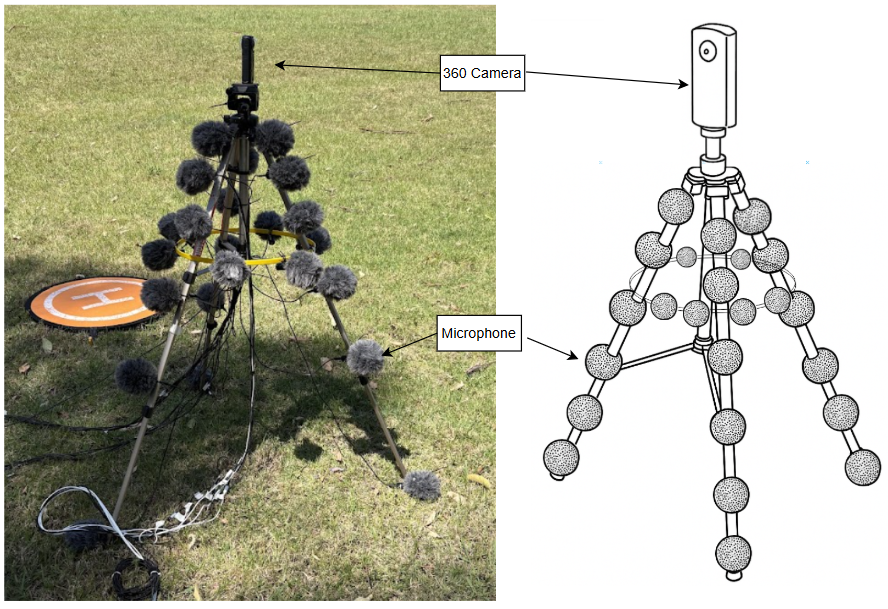}}
  \caption{24 channels microphone array}
  \label{fig:Array}
  \vspace{-0.4cm}
\end{figure}

We assembled a 24-channel microphone array using six Rode microphones mounted on each of the three legs of a standard tripod, forming an upright tetrahedral array as shown in \cref{fig:Array}.

The remaining six microphones were arranged in a horizontal circular "yellow" ring. The array geometry provides progressive inter-microphone distances ranging from 4 cm to 1.1 m, optimizing for spatial aliasing and directional resolution in the frequency band between approximately 200 Hz and 4000 Hz, based on the array aperture and the speed of sound. Four Zoom F6 multichannel recorders (6 channels each) were synchronized via a synthetic impulse: channel 1 of every unit is placed at the array origin, the impulse is played once, and all files are shifted until their peak samples coincide, yielding ±1 sample at 48 kHz (7 mm acoustic error). An Insta360 X4 camera was mounted on the top to capture visual reference of drone takeoff and synchronize audio and flight logs temporally.

\subsection{Field Recordings and Data Acquisition}
A total of six open-field recording sessions were conducted using a DJI Air 3 drone, across three different dates and two distinct locations. Each session produced 24-channel synchronized audio recordings $x_n(t)$ signals in \texttt{.wav} format, 360° video files, and GPS log files from the drone’s onboard telemetry. The GNSS (Global Navigation Satellite System) on the DJI Air 3 provides ±0.5 m position accuracy. An additional log file was recorded with the drone stationary at the array’s location to define the Cartesian origin $(0,0,0)$ for GPS transformation and reference position. This reference enabled accurate computation of azimuth and elevation angles relative to the array.

\subsection{Beamforming Map Construction and Spectral Feature Extraction}

Each GPS log point from the drone is transformed to a Cartesian coordinate system with the microphone array origin at $(0, 0, 0)$. The azimuth ($\phi$) and elevation ($\theta$) are then calculated as:

\setlength{\abovedisplayskip}{4pt} \setlength{\belowdisplayskip}{4pt}
\begin{equation}
    \phi = \arctan2(y, x), \quad \theta = \arcsin\left( \frac{z}{\sqrt{x^2 + y^2 + z^2}} \right)
\end{equation}

For beamforming, the DAS output $y(t)$ in a direction $(\phi, \theta)$ is computed by delaying the signals at each microphone $x_n(t)$ by a steering delay $\tau_n$ and summing, N is the number of microphones:

\begin{equation}
    y(t; \phi, \theta) = \frac{1}{N}\sum_{n=1}^{N} x_n(t - \tau_n(\phi, \theta))
\end{equation}

In general, beamforming relies on the far-field assumption that the sources are far away and the waves become  planar at the array position \cite{bai_acoustic_2013}. Given the position vector of each microphone $\mathbf{p}_n$ in the array, the delay $\tau_n$ is obtained via the projection of $\mathbf{p}_n$ onto the direction vector:

\begin{equation}
\tau_n = \frac{1}{c} \cdot (\mathbf{p}_n \cdot \mathbf{u}_{\phi, \theta}),
\label{eq:delay_projection}
\end{equation}

\begin{equation}
\mathbf{u}_{\phi, \theta} = 
\begin{bmatrix}
\cos\theta \cos\phi \\
\cos\theta \sin\phi \\
\sin\theta
\end{bmatrix}.
\label{eq:unit_vector}
\end{equation}

where $c$ is the speed of sound in air (typically chosen as 343 m/s). The signals are then shifted by these delays, and their average yields the beamformed output for a given steering direction. This approach is repeated for all directions $(\phi, \theta)$ in a discretized spatial grid to construct a full acoustic energy map.
The acoustic beamformed maps were initially computed over a 2D rectangular grid covering azimuth angles in $[-180^{\circ}, 180^{\circ}]$ and elevation angles in $[0^{\circ}, 90^{\circ}]$, using a resolution of $4^{\circ}$ in both dimensions. For each azimuth-elevation pair, a DAS beamforming algorithm was applied to align and sum time-delayed microphone signals over a 100 ms window, sampled at 48~kHz. This yields a time-domain waveform of 4800 samples per spatial direction, producing a 3D tensor snapshot with shape $(A \times E \times T)$, where $A$ is the number of azimuth bins, $E$ the number of elevation bins, and $T$ the number of time samples.

To convert these maps into a spectral representation, each time-
domain waveform is transformed using the FFT. Frequency-domain
features are computed by applying the real-valued FFT directly to each 4800-sample time-domain frame, using a rectangular window and no overlap between consecutive frames. Only the 200–2200 Hz band is
retained, corresponding to the dominant energy of the DJI Air 3
drone, uniformly divided into F = 16 bins, which are globally
normalized across spatial directions per frame. The result is a
tensor of shape (A × E × F ).

To better align with the spherical nature of the acoustic scene and reduce distortion in convolutional layers, the $(A \times E \times F)$ data is reprojected into a polar grid. In this transformation, elevation is mapped to radial distance from the center (with $90^{\circ}$ at the center and $0^{\circ}$ at the outer edge), and azimuth is mapped to angular position around the circle. The resulting spatial layout is a square grid of size $(2E \times 2E)$, where the angular geometry is preserved. The final input tensor $\mathbf{X}$ used for learning has shape $(2E \times 2E \times F)$.

\subsection{Dataset Construction and GPS-Based Labeling}

The dataset was recorded in multiple sessions. The training set comprises 30 minutes of drone flight from March 2025, along with an additional 22 minutes of drone flight from November 2024, plus approximately 5 minutes of ambient noise, all at Site 1. Test 2 consists of 20 minutes of drone flight and approximately 5 minutes of ambient noise from March 2025, also at Site 1. Test 1 includes 3 minutes of drone flight and 1 minute of ambient noise from October 2024, recorded at Site 2, under unseen environmental conditions. Each flight session is treated as an independent data segment to prevent information leakage during training and evaluation.

Synchronized GPS logs from the drone are converted into spherical coordinates relative to the microphone array origin. Each 100~ms frame is annotated using a radial angular threshold $\delta = 10^{\circ}$ around the ground-truth direction of arrival (DOA): all pixels within this threshold are labeled as 1 (drone-present), and the rest as 0 (drone-absent). This creates binary segmentation masks $\mathbf{Y} \in \{0, 1\}^{2E \times 2E}$.

This labeling strategy compensates for beamforming limitations, such as reduced spatial resolution at low frequencies (due to wider beamwidth) and the presence of side lobes at high frequencies. By allowing spatial tolerance, the model is encouraged to learn smoother and physically grounded segmentation masks, consistent with prior studies on acoustic source mapping~\cite{lee_deep_2021}.

\subsection{Dataset Representation}
Each training example is a pair $(\mathbf{X}, \mathbf{Y})$ where the input tensor $\mathbf{X} \in \mathbb{R}^{2E \times 2E \times F}$ encodes the beamformed spectral information in polar coordinates, and the label $\mathbf{Y} \in \{0, 1\}^{2E \times 2E}$ is the corresponding binary segmentation mask.

\subsection{U-Net Architecture and Hyperparameter Optimization}
We implemented a modified U-Net architecture that accepts rectangular or polar input tensors with shape $(2E \times 2E \times F)$ and outputs a binary segmentation mask $\hat{\mathbf{Y}} \in [0, 1]^{2E \times 2E}$. The architecture consists of an encoder with downsampling convolutional blocks, followed by a symmetric decoder with upsampling layers. Skip connections bridge corresponding encoder and decoder levels to preserve any spatial details lost during the downsampling. Optional attention gates~\cite{oktay_attention_2018} can be applied to skip connections, the bottleneck, or both. The number of base filters, depth of the encoder–decoder path, kernel size, and attention configuration are all tunable hyperparameters.

To optimize the model configuration, we used the Optuna framework to search the hyperparameter space, selecting the configuration that minimized loss computed on the Test 2 split (Site 1, matched conditions); Test 1 (Site 2, unseen conditions) remained fully held out throughout training and model selection. We found 16 FFT bins, 64 base filters, depth 3, $3\times3$ kernels, $\text{lr} = 0.005$, and
skip-attention.

 We use task-specific loss functions depending on label sparsity and class structure. For binary drone segmentation we employ the Tversky loss to explicitly control the FP/FN trade-off under severe foreground sparsity. For multiclass TAU Spatial we use a multilabel Dice + BCE-with-logits objective with per-class positive weights computed from training pixels, which stabilizes learning when class frequencies differ markedly.

\subsection{Inference and Evaluation}
At test time, the U-Net outputs are thresholded, and a centroid is computed over active regions in the predicted mask to estimate the DOA. This estimation is obtained as the centroid of the active pixel set
$S=\{(i,j):\hat{Y}_{i,j}=1\}$ in the predicted mask:
\begin{equation}
(\hat{\phi}, \hat{\theta}) = \frac{1}{|S|}\sum_{(i,j)\in S}(\phi_{i,j},\theta_{i,j})
\end{equation}
computed after thresholding (0.5 on the sigmoid output) and a
$3\times3$ erosion, with fallback to the unfiltered mask if erosion
removes all active pixels.
Metrics are computed per 100 ms frame, then averaged over full trajectories to expose range-dependent accuracy. 

\section{RESULTS}
\label{sec:results}

The results were obtained using two test datasets: \textit{Test 2}, collected in March 2025 at the original site, and \textit{Test 1} in October 2024, recorded at a different location and time under unseen conditions using the same DJI Air 3 drone and no-drone scenarios. 

\begin{figure}[h]
    \centering
    \centerline{\includegraphics[width=0.47\textwidth]{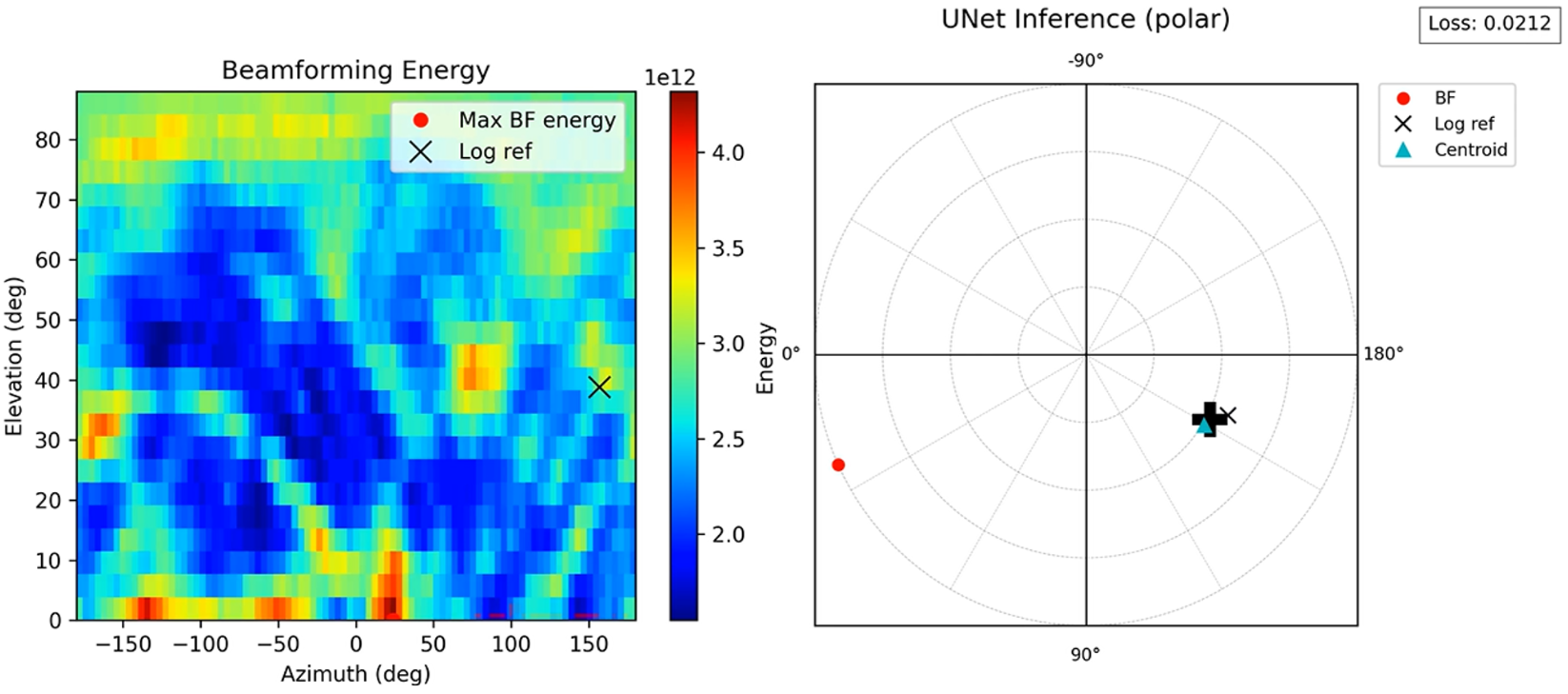}}
    \caption{Instance comparison at 101 m between beamforming localization and U-Net inference.}
    \label{fig:unet_vs_bf}
\end{figure}

On the left side of the \cref{fig:unet_vs_bf}, the beamforming energy map is presented as a function of azimuth and elevation. The black X indicates the ground truth position of the drone (Log ref), extracted from synchronized GPS flight logs, while the red dot shows the location of maximum energy in the beamforming map (BF). It is visually evident that the beamforming-based estimation does not align with the true drone position.

On the right side, the U-Net creates binary segmentation output is displayed in polar coordinates. The black squares represent the segmented region predicted by the model. The blue triangle corresponds to the centroid of the predicted region. It can be observed that the U-Net model provides a significantly more accurate estimation of the drone's position, closely matching the ground truth at a distance at 101 meters between the drone and the microphone array.

\subsection{Performance Analysis by Distance Range}

\begin{figure}[t]
    \centering
    \includegraphics[width=\columnwidth]{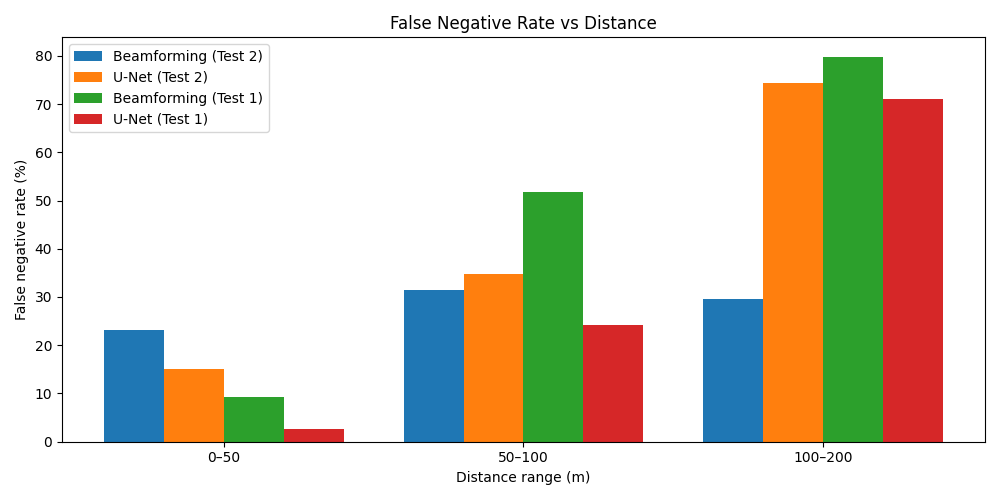}
    \caption{False Negative Rate (FNR) across distance bins for Test 2 and Test 1 datasets.}
    \label{fig:fnr_combined}
\end{figure}

\vspace{0.5em}
\noindent\textbf{False Negative Rate (FNR).} \cref{fig:fnr_combined} shows the FNR as a function of source-array distance, grouped into three bins: 0–50 m, 50–100 m, and 100–200 m. A prediction is classified as a false negative if no output is generated or if the predicted centroid standard deviation exceeds $10^\circ$ over the last 3 observations, indicating high spatial uncertainty. In both datasets, the beamforming approach exhibits higher FNRs across all bins, particularly in the 100–200 m range, where signal energy is weaker. In contrast, the U-Net segmentation model maintains a lower FNR in many cases, suggesting greater robustness to acoustic attenuation and environmental variability. 

\begin{figure}[t]
    \centering
    \includegraphics[width=\columnwidth]{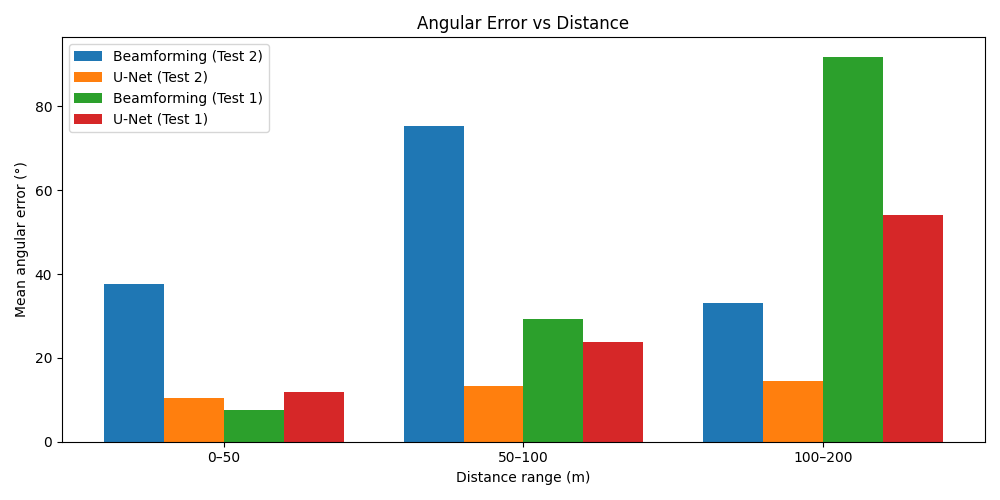}
    \caption{Mean angular error across distance bins for Test 2 and Test 1 datasets.}
    \label{fig:angular_error_combined}
\end{figure}

\vspace{0.5em}
\noindent\textbf{Angular Error.}
\cref{fig:angular_error_combined} reports the mean angular error in degrees for each method and dataset. Beamforming shows increasing error with distance, especially beyond 100 m. The U-Net model consistently outperforms beamforming, with lower average angular errors across all distance bins. Notably, in the 50–100 m bin—where localization tends to be challenging—the model shows stable and accurate predictions even under environmental mismatch in the \textit{Test 1} dataset.

During a five‑minute recording with no drone present, we compared the false‑positive rates. Because there is no target signal in this scenario, performance was evaluated by measuring the standard deviation. Beamforming produced a 67.0\% false‑positive rate, whereas the U‑Net reduced this to just 14.9\%, demonstrating a substantial improvement.

Furthermore, both the false negative rate and the angular localization error can be mitigated by employing multiple synchronized devices with higher microphone density. This is possible within our framework due to the use of low-cost hardware components, enabling scalable deployments. These results confirm that the U-Net-based approach
generalizes better than conventional beamforming across varying test
conditions and distances. 

\begin{figure}[t]
    \centering
    \includegraphics[width=\columnwidth]{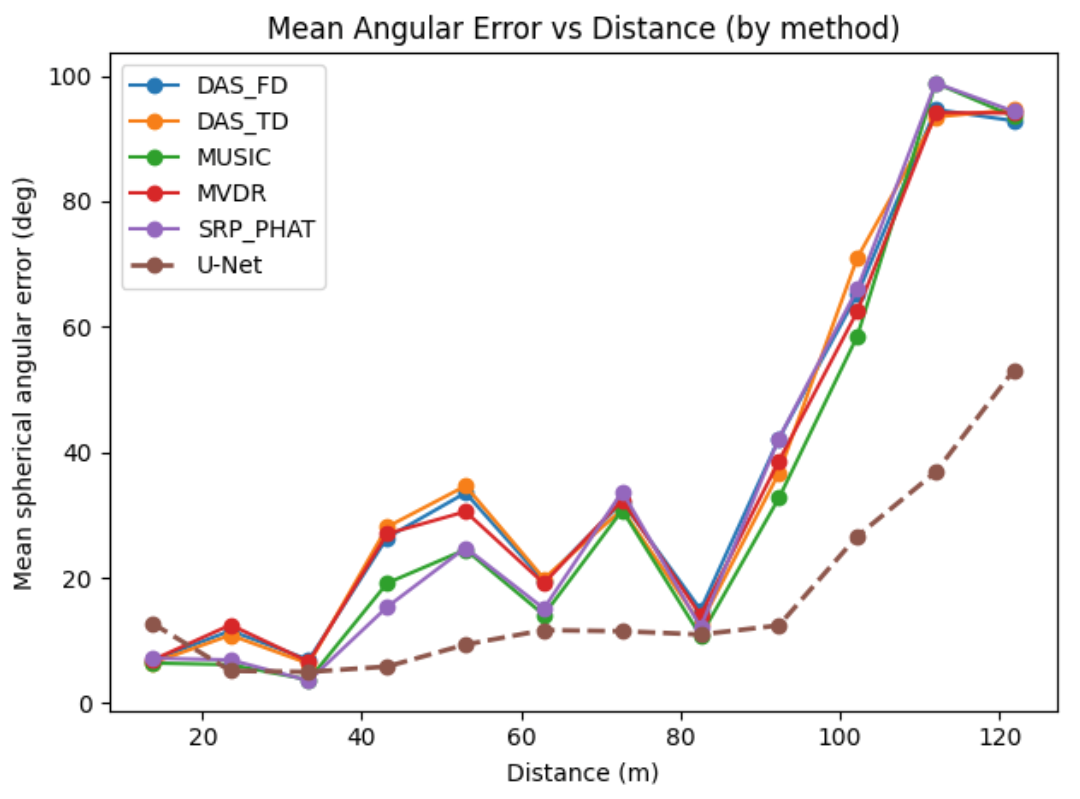}
    \caption{Test~1: mean spherical angular error versus distance for classical SSL baselines (DAS\_FD, DAS\_TD --- frequency and
time-domain implementations of delay-and-sum beamforming,
respectively ---, MUSIC, MVDR, SRP-PHAT) and the proposed U-Net segmentation approach.}
    \label{fig:all_ssl}
\end{figure}

\vspace{0.5em}
\noindent\textbf{Comparison against classical SSL baselines (Test~1).}
\Cref{fig:all_ssl} compares the proposed U-Net against representative classical SSL pipelines on the unseen \textit{Test~1} condition.

For each method, a position estimate is extracted per frame and compared to the GPS-derived ground truth using the spherical angular distance, results are aggregated by distance range.

Across short-to-mid ranges, the classical baselines exhibit comparable trends, but their angular error increases at longer distances (approximately 90 m), consistent with reduced effective SNR and increasing sensitivity to interference and sidelobe structure in the beamformed maps. In contrast, the U-Net maintains substantially lower error over the same distance, indicating that is able to suppress spurious peaks. While the U-Net error also increases with distance, its degradation is markedly slower, supporting the conclusion that learning a spatial mask over the full azimuth-elevation field provides improved robustness compared to selecting a single maximum-response direction.

\subsection{Multiclass Evaluation on the DCASE 2019 TAU Spatial Sound Events Dataset}
\label{subsec:dcase_multiclass}

To further validate the generality of the proposed segmentation-based formulation, we evaluated the model on the DCASE 2019 Task 3 TAU Spatial Sound Events dataset (Task~3)\cite{adavanne_multi-room_2019}, a widely adopted benchmark for multiclass (SELD).

\subsubsection{Dataset and Preprocessing}

The TAU Spatial dataset consists of synthetically generated spatial sound scenes to a tetrahedral four-microphone array. Each recording contains between one and two simultaneously active sound events, positioned at a fixed distance of 1~m and 2~m from the array, with a SNR of approximately 30~dB. The dataset provides time-aligned annotations in azimuth and elevation for multiple sound event classes.

To ensure methodological consistency with the proposed drone pipeline, the dataset was reprocessed into beamformed acoustic maps and converted into a multiclass semantic segmentation task. A regular tetrahedral array geometry was assumed, with an effective radius of 4.2~cm. For each 100~ms audio chunk, delay-and-sum (DAS) beamforming was applied over a discrete azimuth--elevation grid. This produced a three-dimensional beamformed tensor indexed by azimuth, elevation, and time.

Each beamformed time-domain signal was transformed into the frequency domain using the real-valued FFT. We preserve strict parity with the feature extraction pipeline used in the drone experiments. Magnitude spectra were extracted, restricted to the frequency range 100--10,000~Hz, and globally normalized using z-score statistics computed over the training set. The normalized spectra were then grouped into $F$ contiguous frequency bins (we use $F=64$ for TAU Spatial and $F=16$ for UAV). In the TAU Spatial case, we do not apply the polar reprojection used in the UAV pipeline: the elevation annotations are confined to a limited range ($-40^\circ$ to $40^\circ$), so the azimuth–elevation mask remains well-conditioned on the rectangular grid without the severe geometric warping observed in hemispherical maps. In contrast, the UAV experiments span the upper hemisphere ($0^\circ$ to $90^\circ$ elevation), where sampling and label geometry become increasingly distorted near the zenith, motivating the polar (radial-elevation) mapping adopted for the drone setting.

\begin{figure}[t]
    \centering
    \includegraphics[width=\columnwidth]{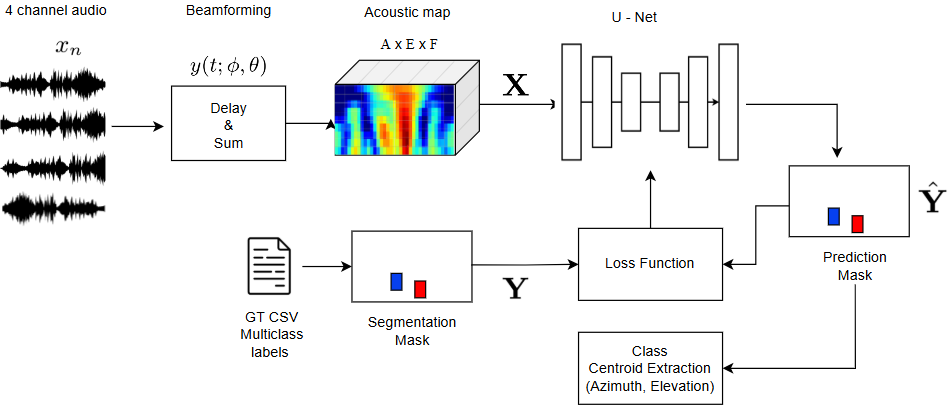}
    \caption{DCASE multiclass pipeline.}
    \label{fig:DCASE}
\end{figure}

To ensure methodological consistency with the proposed drone pipeline, the dataset was reprocessed into DAS beamformed acoustic maps and converted into a multiclass semantic segmentation task, as summarized in Fig.~\ref{fig:DCASE}.

\subsubsection{Multiclass Label Construction}

For each time frame, class-specific binary segmentation masks were generated by comparing the ground-truth event direction with every steering direction in the azimuth--elevation $4^\circ$ grid. Grid points within an angular tolerance of $10^\circ$ from the annotated direction were labeled as active for the corresponding class. This procedure resulted in one binary mask per sound class.

\subsubsection{Network Architecture and Training Protocol}

The same U-Net architecture used in the UAV experiments was employed for the DCASE benchmark, with minor adaptations for multiclass output. The network processes beamformed acoustic maps and predicts a dense segmentation mask with one output channel per sound class.
Training was performed in a multilabel setting using a combined Dice and binary cross-entropy loss. To compensate for severe class imbalance and sparse spatial labels, per-class positive weights were computed from the training set based on the ratio of negative to positive pixels. Model selection and early stopping were driven by the macro-averaged F1 score across classes.

\subsubsection{Results and Discussion}

The segmentation-based U-Net achieves competitive performance on the TAU Spatial dataset despite operating on beamformed energy maps rather than raw multichannel signals. 
\begin{figure}[t]
    \centering
    \includegraphics[width=\columnwidth]{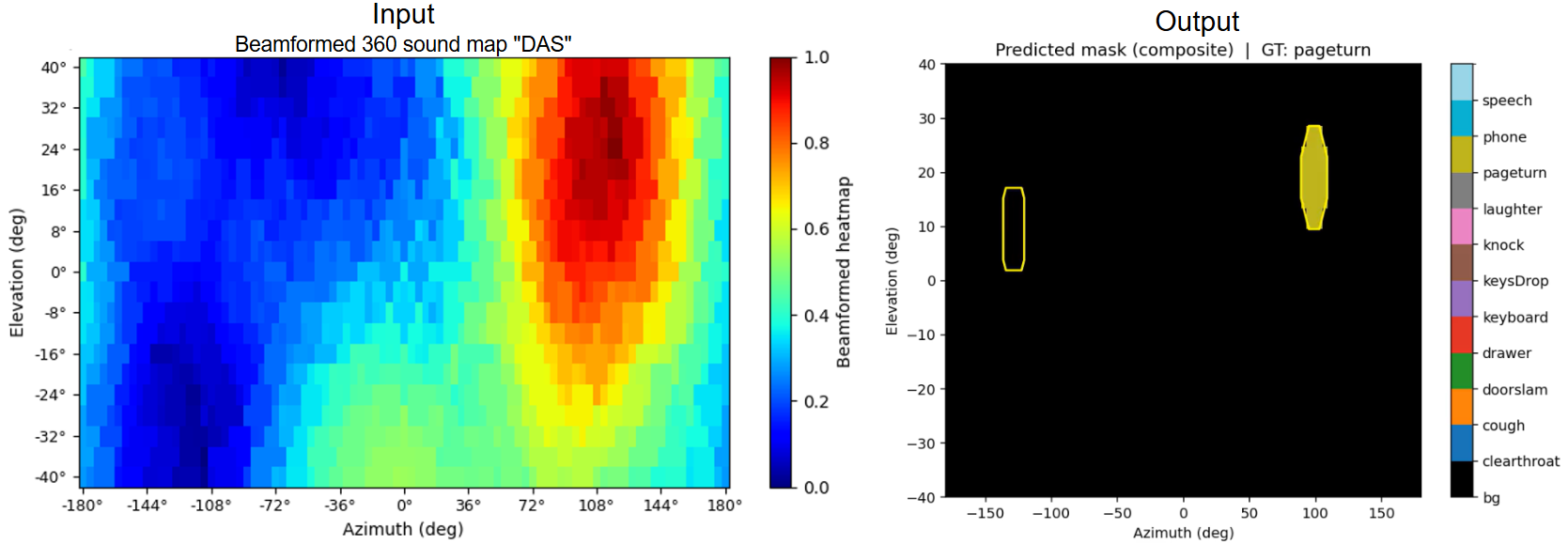}
    \caption{Example on TAU Spatial (DCASE 2019). \textbf{Left:} normalized DAS beamformed spatial map for a single 100\,ms frame (visualized as a 2D azimuth--elevation energy map). \textbf{Right:} U-Net prediction rendered as a composite label map (per-pixel argmax over class probabilities); the ground-truth active class for this frame is \textit{pageturn}.}
    \label{fig:tau19_qual_example}
\end{figure}

Fig.~\ref{fig:tau19_qual_example} shows a qualitative example of the proposed map-to-mask multiclass pipeline on TAU Spatial. Starting from a single-frame DAS beamformed map, the U-Net outputs class-specific regions in the azimuth--elevation grid, effectively transforming a broad and often ambiguous beamformer response into a structured semantic mask. Occasional multiple disconnected detections within the same class can arise due to the limited aperture of the tetrahedral four-microphone array and the sidelobe structure of DAS, which can yield competing spatial hypotheses within a single frame. This observation motivates temporal context (multi-frame input) and/or lightweight spatial association to improve robustness under overlapping events and transient classes.
Table~\ref{tab:tau19_results} situates the proposed pipeline against established SELD baselines on TAU Spatial 2019. On the full evaluation condition (up to two overlapping sources), the segmentation-based U-Net presents an overall F-score of $\approx 68\%$ with a mean DoA error of $4.9^{\circ}$, improving to $79\%$ F-score and $3.0^{\circ}$ DoA error when restricting to frames containing a single active source. In contrast, strong end-to-end SELD systems such as SELDnet and w2v-SELD report $\sim$94--95\% F-score with $\sim$3.7--4.7$^{\circ}$ DoA error, while the DCASE baseline SELDnet exhibits a pronounced degradation in localization error (38.1$^{\circ}$). Overall, these results support that reframing SELD as dense semantic segmentation over spatial acoustic maps can preserve competitive localization accuracy, while the drop under overlapping sources is consistent with the lack of explicit temporal association and multi-instance tracking in a purely frame-wise segmentation formulation. To address this limitation, we are currently developing a temporal variant that ingests multiple consecutive map frames (i.e., increased temporal receptive field) to better capture event dynamics and improve F-score, particularly on TAU Spatial where many classes are impulsive/transient (e.g., knock, drawer, doorslam, keyboard) and thus benefit from temporal context. This differs from the original UAV-driven design target, where the acoustic signature is comparatively sustained and temporally stable, making single-frame segmentation more viable. Finally, we note that the ground-truth mask design should be consistent with the effective array resolution: with fewer microphones the beamformer main lobe is wider (lower angular resolution), which motivates using a larger angular mask tolerance, whereas denser/larger-aperture arrays can support a tighter (smaller) mask due to a narrower main lobe.

\begin{table}[t]
\centering
\caption{TAU Spatial 2019 (DCASE Task 3) results: detection F-score and mean DoA angular error.}
\label{tab:tau19_results}
\begin{tabular}{lcc}
\toprule
\textbf{Model} & \textbf{F1 (\%)} & \textbf{DoA Error ($^\circ$)} \\
\midrule
SELDnet \cite{kapka_sound_2019}               & 94.7  & 3.7  \\
Baseline SELDnet \cite{adavanne_multi-room_2019}                  & 83.1  & 38.1 \\
w2v-SELD \cite{santos_w2v-seld_2023}                         & 94.7 & 4.7  \\
U-Net (eval, up to two sources)   & 68.0  & 4.9  \\
U-Net (one-source subset)         & 79.0  & 3.0  \\
\bottomrule
\end{tabular}
\end{table}

These results demonstrate that reframing SELD as a dense semantic segmentation problem over spatial acoustic maps is not limited to drone acoustics, but generalizes to established multiclass benchmarks. Overall, this experiment confirms that the proposed beamforming-plus-segmentation paradigm provides a possible flexible and array-agnostic alternative to conventional SELD architectures, while maintaining competitive localization accuracy and robust detection behavior across domains.

\section{Conclusion and Future Work}
\label{sec:conclusion_future}

This work presents a U-Net framework for sound source localization (SSL) that reinterprets beamformed acoustic maps as spatial segmentation tasks over the spherical field. By applying U-Net-based convolutional architectures to azimuth--elevation representations of DAS beamformed audio, we address classical limitations in beamforming—such as low-frequency blurring and high-frequency side lobes—thus enhancing angular resolution and robustness. Transforming beamformed maps into polar coordinates further aligns the spatial layout with spherical geometry, reducing distortion and better supporting CNN-based learning, as emphasized in DeepWave~\cite{simeoni_deepwave_2019}.

Unlike end-to-end models requiring raw microphone inputs or fixed array geometries, the proposed U-Net model operates on preprocessed spatial representations, enabling transfer across different microphone array configurations. Combined with real-world drone recordings and GPS-based supervision, the method demonstrates robust generalization across distances and recording conditions, providing a practical solution for low-latency outdoor acoustic perception.

Beyond the UAV scenario, we further validated the proposed beamforming-plus-segmentation paradigm on the DCASE 2019 TAU Spatial Sound Events dataset, a standard benchmark for multiclass SELD. Despite relying on beamformed energy maps rather than raw multichannel waveforms, the segmentation-based formulation achieves competitive direction-of-arrival accuracy when compared to established SELDnet-style architectures. This result confirms that dense spatial mask prediction constitutes a viable alternative parameterization of SELD, in which detection and localization emerge jointly from semantic segmentation of the acoustic field.

Future work will focus on extending the framework to handle spatio-temporal dynamics via recurrent or attention-based models such as Mask R-CNN~\cite{yang_video_2019}, YOLACT++~\cite{bolya_yolact_2022}, or VisTR~\cite{wang_end--end_2021}, enabling explicit temporal association, multi-source instance tracking, and improved performance under overlapping sound events. Additional directions include synthetic data generation for broader acoustic diversity and multi-array fusion for scalable real-time deployments.

By framing SSL and SELD as semantic segmentation tasks on beamformed spatial representations, this work bridges classical array signal processing and modern computer vision, opening new directions for high-resolution, interpretable, and deployable acoustic scene understanding.


\bibliographystyle{IEEEtran}
\bibliography{references}

@misc{rodriguez_uav_2026,
	title = {{UAV} {Acoustic} {Localization} {Dataset}: 24-{Channel} {Beamformed} {Recordings} of a {DJI} {Air} 3 {Drone} with {Synchronized} {GPS} {Flight} {Logs}},
	shorttitle = {{UAV} {Acoustic} {Localization} {Dataset}},
	url = {https://research-data.westernsydney.edu.au/published/1f8cceb0801b11f189a547c836cf5bee},
	doi = {10.26183/07NM-4P35},
	urldate = {2026-07-16},
	publisher = {Western Sydney University},
	author = {Rodriguez, Belman Jahir and Chevtchenko, Sergio and Afshar, Saeed},
	month = jul,
	year = {2026},
	note = {Size: 46 GB, DOI: 10.26183/07nm-4p35},
}

@inproceedings{kapka_sound_2019,
	title = {Sound detection, localization and classification using consecutive ensemble of {CRNN} models},
	language = {en},
	booktitle = {Proceedings of the {Detection} and {Classification} of {Acoustic} {Scenes} and {Events} 2019 {Workshop} ({DCASE2019})},
	author = {Kapka, Sławomir and Lewandowski, Mateusz},
	year = {2019},
}

@article{localization_review_rascon,
	title = {Localization of sound sources in robotics: {A} review},
	volume = {96},
	issn = {09218890},
	shorttitle = {Localization of sound sources in robotics},
	url = {https://linkinghub.elsevier.com/retrieve/pii/S0921889016304742},
	doi = {10.1016/j.robot.2017.07.011},
	language = {en},
	urldate = {2025-01-26},
	journal = {Robotics and Autonomous Systems},
	author = {Rascon, Caleb and Meza, Ivan},
	year = {2017},
	pages = {184--210},
}

@article{jekaterynczuk_survey_2023,
	title = {A {Survey} of {Sound} {Source} {Localization} and {Detection} {Methods} and {Their} {Applications}},
	volume = {24},
	copyright = {https://creativecommons.org/licenses/by/4.0/},
	issn = {1424-8220},
	url = {https://www.mdpi.com/1424-8220/24/1/68},
	doi = {10.3390/s24010068},
	language = {en},
	number = {1},
	urldate = {2025-05-04},
	journal = {Sensors},
	author = {Jekateryńczuk, Gabriel and Piotrowski, Zbigniew},
	month = dec,
	year = {2023},
	pages = {68},
}

@article{knapp_generalized_1976,
	title = {The generalized correlation method for estimation of time delay},
	volume = {24},
	copyright = {https://ieeexplore.ieee.org/Xplorehelp/downloads/license-information/IEEE.html},
	issn = {0096-3518},
	url = {http://ieeexplore.ieee.org/document/1162830/},
	doi = {10.1109/TASSP.1976.1162830},
	language = {en},
	number = {4},
	urldate = {2025-07-07},
	journal = {IEEE Transactions on Acoustics, Speech, and Signal Processing},
	author = {Knapp, C. and Carter, G.},
	month = aug,
	year = {1976},
	pages = {320--327},
}

@article{simeoni_deepwave_2019,
	title = {{DeepWave}: {A} {Recurrent} {Neural}-{Network} for {Real}-{Time} {Acoustic} {Imaging}},
	language = {en},
	journal = {Proceedings of the 33rd International Conference on Neural Information Processing Systems},
	author = {Simeoni, Matthieu and Kashani, Sepand and Hurley, Paul and Vetterli, Martin},
	year = {2019},
}

@phdthesis{dibiase_high-accuracy_2000,
	type = {{PhD} {Thesis}},
	title = {A {High}-{Accuracy}, {Low}-{Latency} {Technique} for {Talker} {Localization} in {Reverberant} {Environments} {Using} {Microphone} {Arrays}},
	language = {en},
	school = {Brown University, Rhode Island},
	author = {DiBiase, Joseph Hector},
	year = {2000},
}

@article{chaudhary_sound_2014,
	title = {Sound {Source} {Localization} using {GCC}-{PHAT} with {TDOA} {Estimation}},
	volume = {1},
	language = {en},
	number = {11},
	journal = {Journal of Basic and Applied Engineering Research},
	author = {Chaudhary, Nitesh Kumar and Verma, Subir and Aditya, Anshu},
	year = {2014},
}

@inproceedings{adavanne_multi-room_2019,
	title = {A {Multi}-room {Reverberant} {Dataset} for {Sound} {Event} {Localization} and {Detection}},
	isbn = {978-0-578-59596-2},
	url = {http://hdl.handle.net/2451/60746},
	doi = {10.33682/1xwd-5v76},
	language = {en},
	urldate = {2025-07-03},
	booktitle = {Proceedings of the {Detection} and {Classification} of {Acoustic} {Scenes} and             {Events} 2019 {Workshop} ({DCASE2019})},
	publisher = {New York University},
	author = {Adavanne, Sharath and Politis, Archontis and Virtanen, Tuomas},
	year = {2019},
	pages = {10--14},
}

@misc{adavanne_direction_2018,
	title = {Direction of arrival estimation for multiple sound sources using convolutional recurrent neural network},
	url = {http://arxiv.org/abs/1710.10059},
	doi = {10.48550/arXiv.1710.10059},
	language = {en},
	urldate = {2025-07-03},
	publisher = {arXiv},
	author = {Adavanne, Sharath and Politis, Archontis and Virtanen, Tuomas},
	month = aug,
	year = {2018},
	note = {arXiv:1710.10059 [cs]},
}

@misc{santos_w2v-seld_2023,
	title = {w2v-{SELD}: {A} {Sound} {Event} {Localization} and {Detection} {Framework} for {Self}-{Supervised} {Spatial} {Audio} {Pre}-{Training}},
	shorttitle = {w2v-{SELD}},
	url = {http://arxiv.org/abs/2312.06907},
	doi = {10.48550/arXiv.2312.06907},
	language = {en},
	urldate = {2025-07-03},
	publisher = {arXiv},
	author = {Santos, Orlem Lima dos and Rosero, Karen and Lotufo, Roberto de Alencar},
	month = dec,
	year = {2023},
	note = {arXiv:2312.06907 [eess]},
}

@article{capon_high-resolution_1969,
	title = {High-resolution frequency-wavenumber spectrum analysis},
	volume = {57},
	copyright = {https://ieeexplore.ieee.org/Xplorehelp/downloads/license-information/IEEE.html},
	issn = {0018-9219},
	url = {http://ieeexplore.ieee.org/document/1449208/},
	doi = {10.1109/PROC.1969.7278},
	language = {en},
	number = {8},
	urldate = {2025-07-02},
	journal = {Proceedings of the IEEE},
	author = {Capon, J.},
	year = {1969},
	pages = {1408--1418},
}

@article{wang_progress_2025,
	title = {Progress in beamforming acoustic imaging based on phased microphone arrays: {Algorithms} and applications},
	volume = {242},
	issn = {02632241},
	shorttitle = {Progress in beamforming acoustic imaging based on phased microphone arrays},
	url = {https://linkinghub.elsevier.com/retrieve/pii/S0263224124019857},
	doi = {10.1016/j.measurement.2024.116100},
	language = {en},
	urldate = {2025-05-07},
	journal = {Measurement},
	author = {Wang, Yong and Deng, Zhi and Zhao, Jiaxi and Kopiev, Victor Feliksovich and Gao, Donglai and Chen, Wen-Li},
	month = jan,
	year = {2025},
	pages = {116100},
}

@inproceedings{ji_transmusic_2024,
	address = {Seoul, Korea, Republic of},
	title = {{TransMUSIC}: {A} {Transformer}-{Aided} {Subspace} {Method} for {DOA} {Estimation} with {Low}-{Resolution} {ADCS}},
	copyright = {https://doi.org/10.15223/policy-029},
	isbn = {9798350344851},
	shorttitle = {{TransMUSIC}},
	url = {https://ieeexplore.ieee.org/document/10447483/},
	doi = {10.1109/ICASSP48485.2024.10447483},
	language = {en},
	urldate = {2025-05-04},
	booktitle = {{ICASSP} 2024 - 2024 {IEEE} {International} {Conference} on {Acoustics}, {Speech} and {Signal} {Processing} ({ICASSP})},
	publisher = {IEEE},
	author = {Ji, Junkai and Mao, Wei and Xi, Feng and Chen, Shengyao},
	month = apr,
	year = {2024},
	pages = {8576--8580},
}

@article{perrot_so_2021,
	title = {So you think you can {DAS}? {A} viewpoint on delay-and-sum beamforming},
	volume = {111},
	issn = {0041624X},
	shorttitle = {So you think you can {DAS}?},
	url = {https://linkinghub.elsevier.com/retrieve/pii/S0041624X20302444},
	doi = {10.1016/j.ultras.2020.106309},
	language = {en},
	urldate = {2025-04-01},
	journal = {Ultrasonics},
	author = {Perrot, Vincent and Polichetti, Maxime and Varray, François and Garcia, Damien},
	month = mar,
	year = {2021},
	pages = {106309},
}

@misc{chen_rethinking_2017,
	title = {Rethinking {Atrous} {Convolution} for {Semantic} {Image} {Segmentation}},
	url = {http://arxiv.org/abs/1706.05587},
	doi = {10.48550/arXiv.1706.05587},
	language = {en},
	urldate = {2025-04-02},
	publisher = {arXiv},
	author = {Chen, Liang-Chieh and Papandreou, George and Schroff, Florian and Adam, Hartwig},
	month = dec,
	year = {2017},
	note = {arXiv:1706.05587 [cs]},
}

@misc{zhou_audio-visual_2023,
	title = {Audio-{Visual} {Segmentation} with {Semantics}},
	url = {http://arxiv.org/abs/2301.13190},
	doi = {10.48550/arXiv.2301.13190},
	language = {en},
	urldate = {2025-04-04},
	publisher = {arXiv},
	author = {Zhou, Jinxing and Shen, Xuyang and Wang, Jianyuan and Zhang, Jiayi and Sun, Weixuan and Zhang, Jing and Birchfield, Stan and Guo, Dan and Kong, Lingpeng and Wang, Meng and Zhong, Yiran},
	month = jan,
	year = {2023},
	note = {arXiv:2301.13190 [cs]},
}

@article{wu_deep_2019,
	title = {Deep {Convolution} {Network} for {Direction} of {Arrival} {Estimation} {With} {Sparse} {Prior}},
	volume = {26},
	copyright = {https://ieeexplore.ieee.org/Xplorehelp/downloads/license-information/IEEE.html},
	issn = {1070-9908, 1558-2361},
	url = {https://ieeexplore.ieee.org/document/8854868/},
	doi = {10.1109/LSP.2019.2945115},
	language = {en},
	number = {11},
	urldate = {2025-05-04},
	journal = {IEEE Signal Processing Letters},
	author = {Wu, Liuli and Liu, Zhang-Meng and Huang, Zhi-Tao},
	month = nov,
	year = {2019},
	pages = {1688--1692},
}

@article{roy_esprit-estimation_1989,
	title = {{ESPRIT}-estimation of signal parameters via rotational invariance techniques},
	volume = {37},
	copyright = {https://ieeexplore.ieee.org/Xplorehelp/downloads/license-information/IEEE.html},
	issn = {00963518},
	url = {http://ieeexplore.ieee.org/document/32276/},
	doi = {10.1109/29.32276},
	language = {en},
	number = {7},
	urldate = {2025-05-04},
	journal = {IEEE Transactions on Acoustics, Speech, and Signal Processing},
	author = {Roy, R. and Kailath, T.},
	month = jul,
	year = {1989},
	pages = {984--995},
}

@article{schmidt_multiple_1986,
	title = {Multiple emitter location and signal parameter estimation},
	volume = {34},
	copyright = {https://ieeexplore.ieee.org/Xplorehelp/downloads/license-information/IEEE.html},
	issn = {0018-926X, 1558-2221},
	url = {https://ieeexplore.ieee.org/document/1143830/},
	doi = {10.1109/TAP.1986.1143830},
	language = {en},
	number = {3},
	urldate = {2025-05-04},
	journal = {IEEE Transactions on Antennas and Propagation},
	author = {Schmidt, R.},
	month = mar,
	year = {1986},
	pages = {276--280},
}

@inproceedings{bertrand_applications_2011,
	address = {Ghent, Belgium},
	title = {Applications and trends in wireless acoustic sensor networks: {A} signal processing perspective},
	isbn = {978-1-4577-1289-0 978-1-4577-1288-3 978-1-4577-1287-6},
	shorttitle = {Applications and trends in wireless acoustic sensor networks},
	url = {http://ieeexplore.ieee.org/document/6101302/},
	doi = {10.1109/SCVT.2011.6101302},
	language = {en},
	urldate = {2025-05-04},
	booktitle = {2011 18th {IEEE} {Symposium} on {Communications} and {Vehicular} {Technology} in the {Benelux} ({SCVT})},
	publisher = {IEEE},
	author = {Bertrand, Alexander},
	month = nov,
	year = {2011},
	pages = {1--6},
}

@misc{adavanne_localization_2019,
	title = {Localization, {Detection} and {Tracking} of {Multiple} {Moving} {Sound} {Sources} with a {Convolutional} {Recurrent} {Neural} {Network}},
	url = {http://arxiv.org/abs/1904.12769},
	doi = {10.48550/arXiv.1904.12769},
	language = {en},
	urldate = {2025-05-04},
	publisher = {arXiv},
	author = {Adavanne, Sharath and Politis, Archontis and Virtanen, Tuomas},
	month = apr,
	year = {2019},
	note = {arXiv:1904.12769 [cs]},
}

@article{tan_sound_2021,
	title = {Sound {Source} {Localization} {Using} a {Convolutional} {Neural} {Network} and {Regression} {Model}},
	volume = {21},
	copyright = {https://creativecommons.org/licenses/by/4.0/},
	issn = {1424-8220},
	url = {https://www.mdpi.com/1424-8220/21/23/8031},
	doi = {10.3390/s21238031},
	language = {en},
	number = {23},
	urldate = {2025-05-04},
	journal = {Sensors},
	author = {Tan, Tan-Hsu and Lin, Yu-Tang and Chang, Yang-Lang and Alkhaleefah, Mohammad},
	month = dec,
	year = {2021},
	pages = {8031},
}

@article{grinstein_steered_2024,
	title = {Steered {Response} {Power} for {Sound} {Source} {Localization}: a tutorial review},
	volume = {2024},
	issn = {1687-4722},
	shorttitle = {Steered {Response} {Power} for {Sound} {Source} {Localization}},
	url = {https://asmp-eurasipjournals.springeropen.com/articles/10.1186/s13636-024-00377-z},
	doi = {10.1186/s13636-024-00377-z},
	language = {en},
	number = {1},
	urldate = {2025-04-28},
	journal = {EURASIP Journal on Audio, Speech, and Music Processing},
	author = {Grinstein, Eric and Tengan, Elisa and Çakmak, Bilgesu and Dietzen, Thomas and Nunes, Leonardo and Van Waterschoot, Toon and Brookes, Mike and Naylor, Patrick A.},
	month = nov,
	year = {2024},
	pages = {59},
}

@misc{oktay_attention_2018,
	title = {Attention {U}-{Net}: {Learning} {Where} to {Look} for the {Pancreas}},
	shorttitle = {Attention {U}-{Net}},
	url = {http://arxiv.org/abs/1804.03999},
	doi = {10.48550/arXiv.1804.03999},
	language = {en},
	urldate = {2025-04-18},
	publisher = {arXiv},
	author = {Oktay, Ozan and Schlemper, Jo and Folgoc, Loic Le and Lee, Matthew and Heinrich, Mattias and Misawa, Kazunari and Mori, Kensaku and McDonagh, Steven and Hammerla, Nils Y. and Kainz, Bernhard and Glocker, Ben and Rueckert, Daniel},
	month = may,
	year = {2018},
	note = {arXiv:1804.03999 [cs]},
}

@article{yalta_sound_2017,
	title = {Sound {Source} {Localization} {Using} {Deep} {Learning} {Models}},
	volume = {29},
	issn = {1883-8049, 0915-3942},
	url = {https://www.fujipress.jp/jrm/rb/robot002900010037},
	doi = {10.20965/jrm.2017.p0037},
	language = {en},
	number = {1},
	urldate = {2025-04-09},
	journal = {Journal of Robotics and Mechatronics},
	author = {Yalta, Nelson and Nakadai, Kazuhiro and Ogata, Tetsuya and {Intermedia Art and Science Department, Waseda University} and {Honda Research Institute Japan Co., Ltd.}},
	month = feb,
	year = {2017},
	pages = {37--48},
}

@article{A_survey_Grumiaux,
	title = {A survey of sound source localization with deep learning methods},
	volume = {152},
	issn = {0001-4966, 1520-8524},
	url = {https://pubs.aip.org/jasa/article/152/1/107/2838290/A-survey-of-sound-source-localization-with-deep},
	doi = {10.1121/10.0011809},
	language = {en},
	number = {1},
	urldate = {2025-03-27},
	journal = {The Journal of the Acoustical Society of America},
	author = {Grumiaux, Pierre-Amaury and Kitić, Srđan and Girin, Laurent and Guérin, Alexandre},
	month = jul,
	year = {2022},
	pages = {107--151},
}

@misc{yang_video_2019,
	title = {Video {Instance} {Segmentation}},
	url = {http://arxiv.org/abs/1905.04804},
	doi = {10.48550/arXiv.1905.04804},
	language = {en},
	urldate = {2025-04-02},
	publisher = {arXiv},
	author = {Yang, Linjie and Fan, Yuchen and Xu, Ning},
	month = aug,
	year = {2019},
	note = {arXiv:1905.04804 [cs]},
}

@misc{wang_end--end_2021,
	title = {End-to-{End} {Video} {Instance} {Segmentation} with {Transformers}},
	url = {http://arxiv.org/abs/2011.14503},
	doi = {10.48550/arXiv.2011.14503},
	language = {en},
	urldate = {2025-04-02},
	publisher = {arXiv},
	author = {Wang, Yuqing and Xu, Zhaoliang and Wang, Xinlong and Shen, Chunhua and Cheng, Baoshan and Shen, Hao and Xia, Huaxia},
	month = oct,
	year = {2021},
	note = {arXiv:2011.14503 [cs]},
}

@misc{redmon_you_2016,
	title = {You {Only} {Look} {Once}: {Unified}, {Real}-{Time} {Object} {Detection}},
	shorttitle = {You {Only} {Look} {Once}},
	url = {http://arxiv.org/abs/1506.02640},
	doi = {10.48550/arXiv.1506.02640},
	language = {en},
	urldate = {2025-04-02},
	publisher = {arXiv},
	author = {Redmon, Joseph and Divvala, Santosh and Girshick, Ross and Farhadi, Ali},
	month = may,
	year = {2016},
	note = {arXiv:1506.02640 [cs]},
}

@misc{ronneberger_u-net_2015,
	title = {U-{Net}: {Convolutional} {Networks} for {Biomedical} {Image} {Segmentation}},
	shorttitle = {U-{Net}},
	url = {http://arxiv.org/abs/1505.04597},
	doi = {10.48550/arXiv.1505.04597},
	language = {en},
	urldate = {2025-04-02},
	publisher = {arXiv},
	author = {Ronneberger, Olaf and Fischer, Philipp and Brox, Thomas},
	month = may,
	year = {2015},
	note = {arXiv:1505.04597 [cs]},
}

@article{bolya_yolact_2022,
	title = {{YOLACT}++: {Better} {Real}-time {Instance} {Segmentation}},
	volume = {44},
	issn = {0162-8828, 2160-9292, 1939-3539},
	shorttitle = {{YOLACT}++},
	url = {http://arxiv.org/abs/1912.06218},
	doi = {10.1109/TPAMI.2020.3014297},
	language = {en},
	number = {2},
	urldate = {2025-04-02},
	journal = {IEEE Transactions on Pattern Analysis and Machine Intelligence},
	author = {Bolya, Daniel and Zhou, Chong and Xiao, Fanyi and Lee, Yong Jae},
	month = feb,
	year = {2022},
	note = {arXiv:1912.06218 [cs]},
	pages = {1108--1121},
}

@article{lee_deep_2021,
	title = {Deep learning-based method for multiple sound source localization with high resolution and accuracy},
	volume = {161},
	issn = {08883270},
	url = {https://linkinghub.elsevier.com/retrieve/pii/S088832702100354X},
	doi = {10.1016/j.ymssp.2021.107959},
	language = {en},
	urldate = {2025-03-27},
	journal = {Mechanical Systems and Signal Processing},
	author = {Lee, Soo Young and Chang, Jiho and Lee, Seungchul},
	month = dec,
	year = {2021},
	pages = {107959},
}

@article{datta_dense-u-net_2024,
	title = {Dense-{U}-{Net} assisted improved audio-visual source tracking for speech enhancement},
	volume = {2023},
	issn = {2732-4494},
	url = {http://digital-library.theiet.org/doi/10.1049/icp.2023.3233},
	doi = {10.1049/icp.2023.3233},
	language = {en},
	number = {35},
	urldate = {2025-03-27},
	journal = {IET Conference Proceedings},
	author = {Datta, J.},
	month = jan,
	year = {2024},
	pages = {128--129},
}

@article{van_veen_beamforming_1988,
	title = {Beamforming: a versatile approach to spatial filtering},
	volume = {5},
	copyright = {https://ieeexplore.ieee.org/Xplorehelp/downloads/license-information/IEEE.html},
	issn = {0740-7467},
	shorttitle = {Beamforming},
	url = {http://ieeexplore.ieee.org/document/665/},
	doi = {10.1109/53.665},
	language = {en},
	number = {2},
	urldate = {2025-03-21},
	journal = {IEEE ASSP Magazine},
	author = {Van Veen, B.D. and Buckley, K.M.},
	month = apr,
	year = {1988},
	pages = {4--24},
}

@book{bai_acoustic_2013,
	edition = {1},
	title = {Acoustic {Array} {Systems}: {Theory}, {Implementation}, and {Application}},
	copyright = {http://doi.wiley.com/10.1002/tdm\_license\_1.1},
	isbn = {978-0-470-82723-9 978-0-470-82725-3},
	shorttitle = {Acoustic {Array} {Systems}},
	url = {https://onlinelibrary.wiley.com/doi/book/10.1002/9780470827253},
	doi = {10.1002/9780470827253},
	language = {en},
	urldate = {2025-02-24},
	publisher = {Wiley},
	author = {Bai, Mingsian R. and Ih, Jeong‐Guon and Benesty, Jacob},
	month = jan,
	year = {2013},
}

@article{martinez-carranza_review_2020,
	title = {A {Review} on {Auditory} {Perception} for {Unmanned} {Aerial} {Vehicles}},
	volume = {20},
	copyright = {https://creativecommons.org/licenses/by/4.0/},
	issn = {1424-8220},
	url = {https://www.mdpi.com/1424-8220/20/24/7276},
	doi = {10.3390/s20247276},
	language = {en},
	number = {24},
	urldate = {2025-01-26},
	journal = {Sensors},
	author = {Martinez-Carranza, Jose and Rascon, Caleb},
	month = dec,
	year = {2020},
	pages = {7276},
}







\end{document}